\documentclass[12pt,preprint]{aastex}

 \def\ltsima{$\; \buildrel < \over \sim
\;$} \def\simlt{\lower.5ex\hbox{\ltsima}}            
\def\gtsima{$\; \buildrel > \over \sim \;$}
\def\simgt{\lower.5ex\hbox{\gtsima}}            

\def\lfir{{L$_{\rm FIR}$\/}}
\def\lsol{{L$_\odot$\/}}
\def\lir{{L$_{\rm IR}$\/}}
\def\m2{{M$_{\rm H_2}$\/}}

\received{2001 October 18}
\begin{document}

\title{\sc The Circum-Galactic Environment of Bright IRAS Galaxies }

\author{Y.  Krongold and D. Dultzin-Hacyan}

\affil{Instituto de Astronom\'\i a, UNAM, Aptdo.  Postal 70-264,
M\'exico, D.  F.  04510, M\'exico\\e-mail:yair@astroscu.unam.mx,
deborah@astroscu.unam.mx}

\and

\author{P.  Marziani} \affil{Osservatorio Astronomico, Vicolo dell'
Osservatorio 5, I--35122, Padova, Italia\\ e-mail:
marziani@pd.astro.it}

\begin{abstract}

This paper  studies systematically, for the first time, the circumgalactic environment of bright IRAS
galaxies as defined by \citet{s89}. While the role of gravitational interaction for luminous and
ultraluminous IRAS galaxies has been well established by various studies, the situation is by far more
obscure in the IR luminosity range of the bright IRAS sample, $10^{10}$\lsol $\simlt$ \lfir\ $\simlt
10^{11}$ \lsol. To easily identify nearby companion galaxies, the bright IRAS sample was restricted to 87
objects with redshift range $0.008 \leq z \leq 0.018$ and galactic latitude $\delta \geqq \mid
30^{o}\mid$. A control sample, selected from the Center for Astrophysics redshift survey catalogue,
includes 90 objects matching the Bright IRAS sample for distribution of isophotal diameter, redshift, and
morphological type. From a search of nearby companion galaxies within 250 Kpc on the second-generation
Digitized Sky Survey (DSS-II), we found that the circumgalactic environment of the Bright IRAS galaxies
contains more large companions than the galaxies in the optically selected control sample, and is similar
to that of Seyfert 2 galaxies. We found a weak correlation over a wide range of far IR luminosity  (10$^9$
\lsol $\simlt$ \lfir $\simlt 10^{12.5}$ \lsol) between projected separation and \lfir\, which confirms a
very close relationship between star formation rate of a galaxy and the strength of gravitational
perturbations. We also find that the far IR colors depend on whether a source is isolated or interacting.
Finally, we discuss the intrinsic difference and evolution expectations for the bright IRAS galaxies and
the control sample, as well as the relationship between starbursting and active galaxies.

\end{abstract}


\section{Introduction}

The IRAS Bright Galaxy Survey (hereafter BIRG survey) by
\citet{s89} and by \citet{s95} (southern extension) includes all
the galaxies brighter than $5.4$ Jy at $60 \mu$m. The IRAS Bright
galaxies are therefore, by definition, the brightest
extragalactic objects in the sky at $60 \mu$m. From this survey
we learned a wealth of astrophysical information: (1) the far IR
(FIR) emission dominates the total luminosity in a significant
fraction of galaxies; (2) at luminosity $\log(L_{ir}/ L_\odot)
\simgt 11$ (the so-called Luminous Infrared Galaxies, LIRGs), IR
selected galaxies become more numerous than optically selected
Starburst and Seyfert galaxies of comparable bolometric
luminosity. At luminosity $\log(L_{ir}/ L_\odot) \simgt 12$, the
so-called Ultra-Luminous Infrared Galaxies (ULIRGs) exceed the
space densities of QSOs by a factor of 1.5-2 (\citet{sm96};
\citet{ssi99}).

A considerable number of studies suggest a strong relation between galaxy interactions and the highest IR
luminosity. ULIRGs are often found to be interacting/merging systems (\citet{ssi99}). However, the
environment of moderately luminous infrared galaxies ($10^{10}L_{\odot} \leq L_{fir} \leq
10^{11}L_{\odot}$, hereafter MIRGs), and luminous infrared galaxies is not well known yet. In this paper,
we study the circumgalactic environment of 87 galaxies from the BIRG Survey, with luminosity range
$10^{10}$\lsol $\leq $\lfir $\leq 10^{12}$ \lsol. The sample is composed by MIRGs and a few LIRGs. We also
consider whether a correlation may be present between FIR properties and the projected separation of the
BIRG and its nearest companion (\S \ref{corr}). We then compare the BIRG environment to the one of Seyfert
1 and 2 galaxies (\S \ref{seyf}). Finally, we discuss the implications of interaction-induced \lfir\
enhancement for secular evolution of galaxies and for the relationship between starbursting and active
galaxies (\S \ref{disc}). In the following discussion, we adopt a Hubble constant of $H_o =
75~Km~s^{-1}~Mpc^{-1}$.

\section{Sample Selection}

\subsection{Bright IRAS Sample}

The Bright IRAS sample consists of 87 objects, and was compiled from the IRAS Bright Galaxy Survey by
\citet{s89} for the Northern hemisphere, and by \citet{s95} for the Southern one. All objects with
galactic latitude $|b_{II}| \geq 30^{\circ}$\ were included in the sample. In this way we avoid sampling
the galactic plane, where a bias in the detection of companions is expected because of both absorption and
crowding. We further restrict our selection to a volume-limited sample (redshift range $0.008 \leq z \leq
0.018$). A $V/V_{max}$\ test \citep{schmidt68} gives a value of 0.47 $\pm$ 0.05 (rms). Since the BIRG
Survey is highly complete, this sample is expected to be also complete.

The lower z limit (0.008) was chosen to avoid objects with very large angular size, while the upper z
limit (0.018) was set to include the largest possible number of objects and at the same time to avoid very
small angular sizes, especially for the companions that could be confused with stars (see \S\
\ref{ident}). It is important to point out that all the objects selected with the former restrictions lie
in the luminosity range $10^{10}L_{\odot} \leq L_{fir} \leq 10^{12}L_{\odot}$, with MIRGs being the wide
majority ($\approx 92\%$ are MIRGs, and 8\% are LIRGs). The $60\mu$m luminosity (in
$ergs~s^{-1}~\AA^{-1}$) distribution and the \lfir\ (in solar units) distribution of the BIRG sample are
shown in the left and right panels of Fig. \ref{Fig01} respectively.


\subsection{Control Samples}

\begin{figure}

\plotone{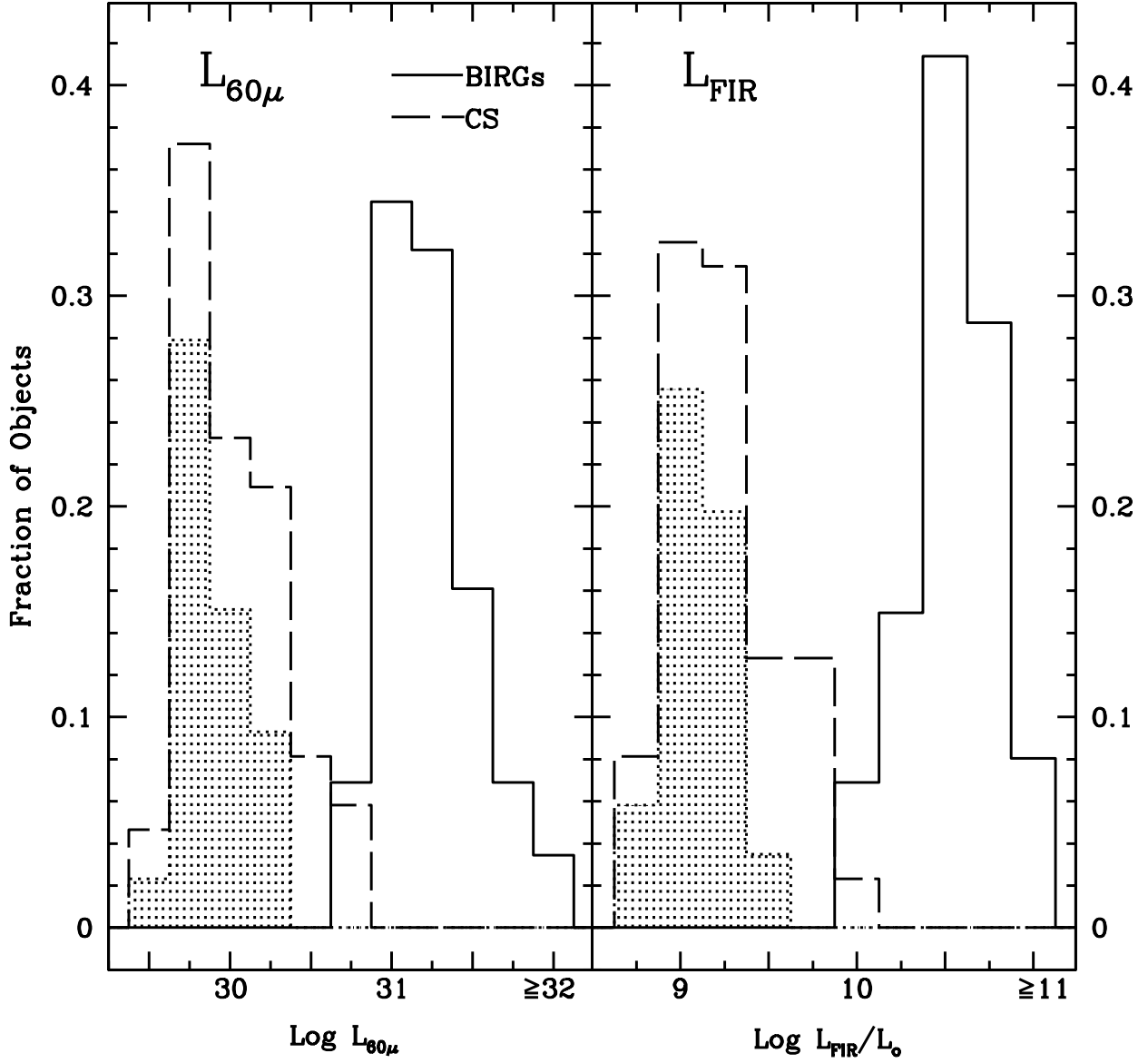}

\figcaption[f1.eps]{Luminosity at 60 $\mu$m (in $ergs~s^{-1}~\AA^{-1}$) (left panel) and \lfir\ (in solar
units) (right panel) for BIRGs and CS galaxies. The solid line traces the distribution of  the BIRG
sample, and the dashed one that of the control sample, including detections as well as upper limits. The
filled area identifies the distribution of objects from the CS whose specific luminosity at 60 $\mu$m and
\lfir\ are known only as upper limits.
\label{Fig01}}
\end{figure}

The control sample for this study was randomly extracted from a list of more than 10,000 objects of the
CfA Catalog \citep{huchra93}. This control sample matches: (1) the isophotal diameter, (2) the redshift,
and (3)the  Hubble morphological type distribution of the BIRG sample. Only objects with galactic latitude
$|b_{II}| \geq 30^{\circ}$\ were included. The control sample consists of 90 objects. A $V/V_{max}$\ test
\citep{schmidt68} gives a value of 0.48 $\pm$ 0.04 (rms). The CS objects are low infrared emitters as can
be seen in Fig. \ref{Fig01}. Their flux at $60 \mu$ is usually much smaller than 5.4 Jy, and their
luminosity at this wavelength is systematically smaller than the luminosity of the BIRGs. Objects without
a detection are treated as upper limits, using the flux density limits of IRAS. The distribution of upper
limits is shown by the filled histograms in Fig. \ref{Fig01}. The absolute B magnitude distribution was
not matched. The B luminosity may be partially correlated with the IR luminosity, since both are enhanced
through star formation processes. Therefore, any attempt to match the B luminosity could bias the control
sample towards galaxies with high infrared luminosity, which is what we want to avoid.

\section{Analysis}

\subsection{Identification of Galaxy Companions \label{ident}}

As in previous environmental studies (\citet{k01},
\citet{dd99a}), the search for galaxy companions was performed
automatically on the DSS-II with the latest version (1998) of
FOCAS (Faint Object Classification and Analysis System;
\citet{jarvis81}), and was limited to galaxy companions that
could be unambiguously distinguished from stars by the FOCAS
algorithm. Each set of pixels with a flux value larger than the
sky threshold is considered an object by FOCAS, and can be
classified as galaxy or star only if its diameter is larger than
4 pixels. Since the scale of the DSS-II plates is $\approx$ 1.0
arcsec per pixel, the minimum angular size to which FOCAS is able
to classify objects on the DSS-II is $\approx$ 4 arcsec (which
corresponds to $\approx$ 1.4 Kpc of projected linear distance).
However, we further restrict our search to companion galaxies of
diameters $D_C \geq 5$ Kpc. With our methodology we cannot study
smaller objects because the distribution of companions is
dominated by optical pairs (not physically associated; as pointed
out in \S 4.1, optical pairs are the wide majority also in the
case of companion diameter between 5 and 10 Kpc). A third
limitation is that FOCAS classifies bright stars as galaxies,
since they appear as extended objects due to scintillation
effects. To avoid gross mis-classifications, we checked by eye on
the computer screen each object classified by FOCAS as a galaxy.
Furthermore, border-line objects of marginally resolved
appearance were not taken into account to avoid also second order
mis-classifications. Effect of plate quality, point spread
function, sky background, and of automatic identification and
measurement of companion and background galaxies have been
discussed in \citet{k01}. They will not be discussed again here;
the same effects are still influencing the analysis of the DSS-II.

As is customary in many previous works (e.g. \citet{dd99b}; \citet{k01}), the fraction of objects with
``physical" companions $f_{phys}$\ is taken as the fraction with one or more observed companions
$f_{obs}$, reduced by the fraction of galaxies with one or more optical companions (derived from Poisson
distribution), namely $f_{phys} = f_{obs} - f_{opt}.$\ The number of background galaxies expected to
follow a Poisson statistics has been obtained as described by \citet{k01}.

\section{Results}

\subsection{Companions Within 3$D_S$}

We looked  for companions  in a circular area with radius equal to 3
times the diameter of the central object ($3D_S$). Our results are
summarized in Table 1.

\paragraph{Companion  diameter $10Kpc \geq D_C \geq 5Kpc$}
Of 87 BIRGs galaxies, $\approx$ 40{\%} has at least one companion within 3$D_S$, vs. 43 {\%} of the 90
objects of the CS. The expected number of optical companions from Poisson statistics is 36 {\%}, and 36
{\%} for the BIRG, and CS, respectively. If optical companions are subtracted, $f_{phys}$\ is $\approx$
4{\%}, and 6.5{\%} for the BIRG and the CS, respectively. These results show that there is not a
significant excess of companions between Bright IRAS galaxies with respect to non-active galaxies, if all
companion galaxies with  $5 Kpc \simlt D_C \simlt 10 Kpc$\ are taken into account. However, this result
should be viewed with caution since $f_{opt} \gg f_{phys}$. A statistical approach is not appropriate in
this companion size range. Any inter-sample difference can be proved as significant only if $f_{phys}$\ is
estimated from redshift measurements for all companion galaxies.

\paragraph{Companion diameter $D_C \geq 10Kpc$} Of 87  BIRG galaxies,
$\approx$ 58.4{\%} has at least one companion of diameter $D \geq
10Kpc$\ within a search radius 3 $D_S$, against $\approx$29 {\%}
of the 90 objects of the CS.  The expected number of optical
companions from Poisson statistics is 20{\%}, and 18.4{\%} for
the BIRG and CS respectively. If $f_{opt}$\ is subtracted,
$f_{phys}$ is $\approx$ 38.4{\%} and 10.9{\%} for the BIRG and
the CS, respectively. These results show an excess of large
companions ($D_C \geq 10 Kpc$) in the Bright IRAS galaxies with
respect to non-active galaxies. A $\chi^2$ test gives a
confidence level for this result of 99.9{\%}.

\subsection{Cumulative Distribution of the  Nearest  Companion in the
BIRG Sample and in the Control Sample}

\begin{figure}

\plotone{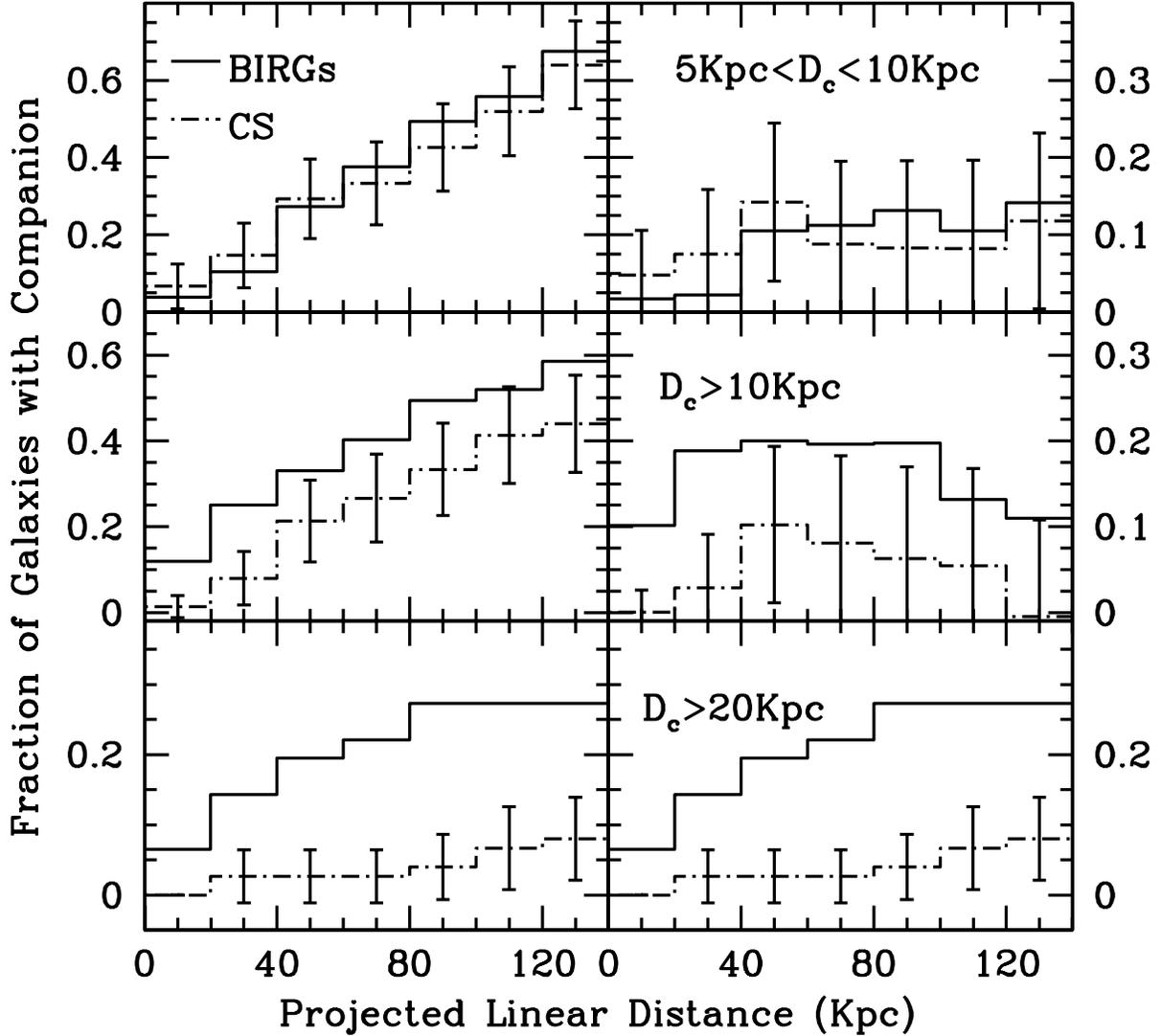}

\figcaption[f2.eps]{Left: Cumulative distributions of nearest observed companion to BIRG galaxies binned
over 20 Kpc, with a projected linear distance limit of 140 Kpc. Right: Distributions of ``physical"
companions (corrected for optical companions with Poisson statistics). The upper panels show the
distributions for galaxies with diameter 5 Kpc $\leq$ $D_c$ $\leq$ 10 Kpc, the middle panels show
``bright" companion galaxies whose diameters are $D_c \geq 10$ Kpc, and the lower panels show companions
with $D_c \geq$ 20 Kpc. The solid line corresponds to the BIRG sample, while the dotted-dashed  refer to
 the control sample. The error bars on the CS are at
a $2\sigma$ confidence level. \label{fig02}}
\end{figure}

The search radius {\em in all the cases} was taken as 250 Kpc of projected linear distance, beyond which
we assumed a ``non detection." The left hand side of Fig. \ref{fig02} presents three panels with the
cumulative distribution of the nearest companion (without correction for optical companions) up to a
projected linear distance ($d_p$) of 140 Kpc. The upper panel shows the cumulative distribution of
companions with diameter in the range $5 Kpc \simgt D \simgt 10 Kpc$, {\em without subtraction of optical
companions}. The middle panel shows the cumulative distribution of companions with diameter $D\geq 10Kpc$,
and the lower panel shows the same distribution for companions with $D\geq$ 20 Kpc. The error bars on the
control samples frequencies were set with a ``bootstrap" technique \citep{boot} by randomly re-sampling
the control sample galaxies into a large number (3000) of pseudo-control samples (i. e., we built 3000
pseudo-control samples of 90 randomly-selected galaxies). The uncertainty on the companion frequency was
set as equal to twice the standard deviation measured from the distribution of 3000 companion frequencies
computed for each pseudo control sample. Comparing the environments of BIRG galaxies and control sample
galaxies, it is found that there is a statistically significant excess of bright companions ($D_C \simgt
10 Kpc$) in the infrared emitters. For companion diameters $5 Kpc \leq D_C \leq 10 Kpc$ the samples show
no significant difference.

\subsection{Distribution of Objects with a ``Physical" Companion}

From the Poisson statistics, we calculated $f_{opt}$\ at distances 20 Kpc, 40 Kpc, etc. By subtracting
this number from $f_{obs}$, we built the distribution of the nearest physical companion. The right hand
side of Fig. \ref{fig02} presents three panels with this distribution, up to $d_p \approx$ 140 Kpc. The
upper panel shows the  distribution of ``physical" companions with diameter in the range $5kpc \leq D_C
\leq 10 Kpc$. The middle panel shows the distribution of physical companions with diameter $D_C \geq 10
Kpc$. The lower panel shows the  distribution of physical companions with $D\geq 20 Kpc$. In the latter
case, the surface density of objects above this diameter (20 Kpc) is very low, and the probability of
finding optical companions is negligible. Therefore, the cumulative $f_{obs}$ is $ \approx f_{phys}$. The
error bars on the CS  frequencies were again set with the ``bootstrap" technique. As before, the results
show a statistically significant excess of bright ``physical" companions ($D_C \geq 10 Kpc$) in the BIRG
galaxies. For companion diameters $\leq 10 Kpc$, there is no significant difference between the two
samples.

\subsection{BIRGs vs. Sy1s and Sy2s}

\begin{figure}

\plotone{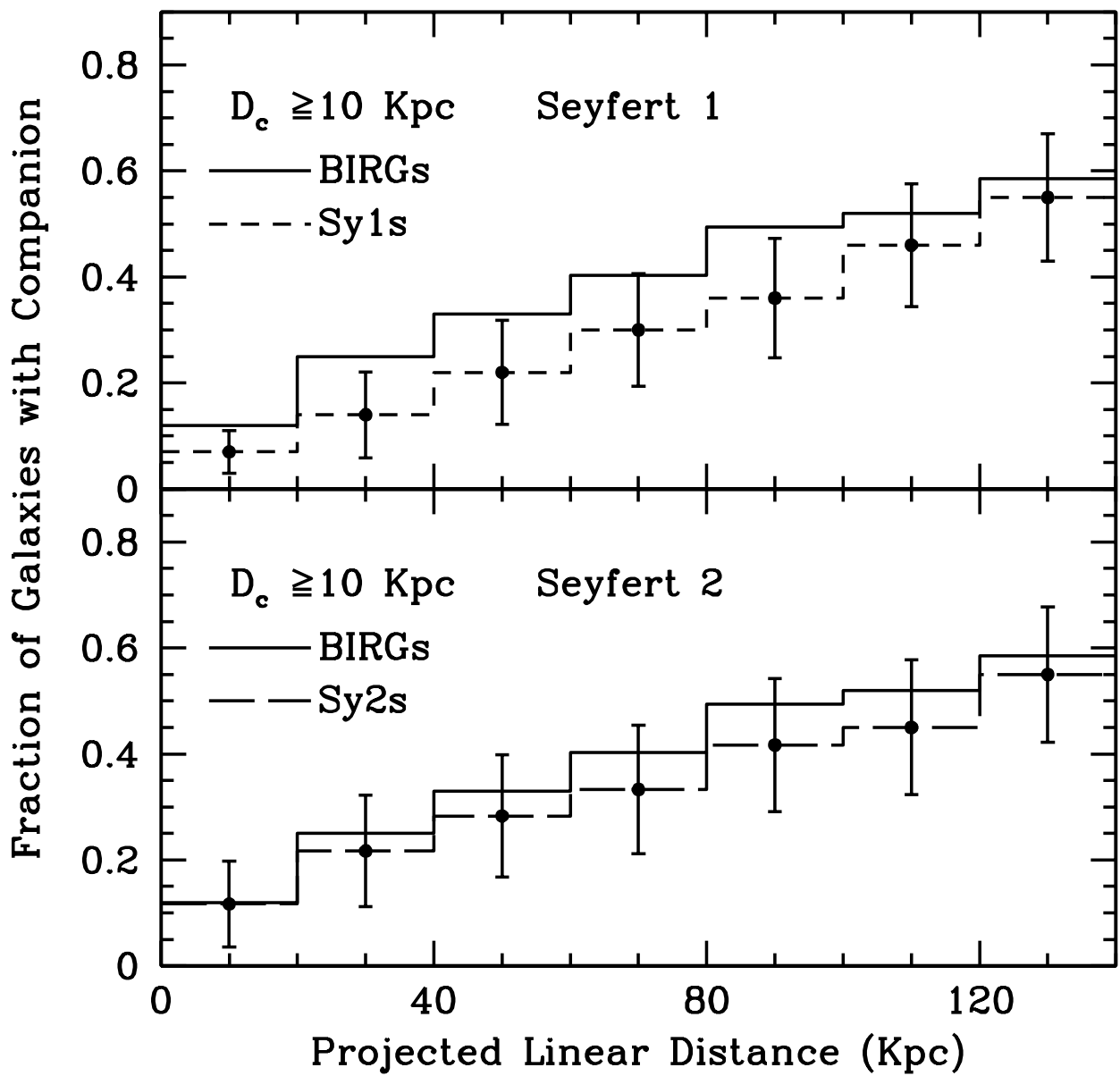}

\figcaption[f3.eps]{ The distributions of the nearest companion with diameter $D_c \geq 10$ Kpc, binned
over 20 Kpc, up to a projected linear distance of 140 Kpc, for Sy1, Sy2, and BIRG galaxies. Upper panel:
BIRG vs. Sy1 galaxies. Lower panel: BIRG vs. Sy2 galaxies. The solid line corresponds to the BIRG sample,
while the dashed refers to Sy1 in the upper panel and  to Sy2 in the lower one. The error bars are set at
a $2\sigma$ confidence level. \label{fig03}}

\end{figure}

In order to study the difference between  the environment of BIRGs, Seyfert 1, and Seyfert 2 galaxies,  we
used the data obtained for Seyfert environments by \citet{dd99a}. The comparison is straightforward since
z range of our BIRG sample, search radius and diameter limits are identical to the ones of the Seyfert 2
sample of \citet{dd99a}. The cumulative distribution for the projected distance $d_P$\ of the first
observed companion for these objects is presented in Fig. \ref{fig03}. The error bars in  Fig. \ref{fig03}
were set with the bootstrap technique, and are at a 2$\sigma$\ confidence level, as in \citet{dd99a}. The
lower panel of Fig. \ref{fig03} shows that there is almost no difference in the distribution of first
companion distances between the BIRG galaxies and the Sy2s. On the contrary, the upper panel shows that
there is a statistically significant excess (a $\chi^2$ test gives a confidence level of 99\%) of
companions in the BIRG sample  with respect to Sy1 galaxies. A similar difference was found between Sy1s
and Sy2s \citep{dd99a}.

\subsection{Group Membership}

We searched in the environment of our objects to determine
whether they  belong to an association of galaxies. We considered
any object with at least two companions with diameter $D_C\geq 10
Kpc$ within a circle of radius $200$ Kpc as member  of a group of
galaxies. 25\%\ of the 87 galaxies from the BIRG sample matched
the former criteria (43\%\ of the BIRG galaxies with at least one
companion with $D_C \simgt 10 Kpc$). Only 4.3\% of the control
sample objects were members of groups as defined here (10\% of
the CS galaxies in pairs with $D_C \simgt$ 10 Kpc). The results
imply that BIRG galaxies are more frequently found as members of
groups than IR-low emission galaxies.

Only $\approx 14\%$\ of the objects in groups belong to compact
groups of the Hickson Catalog \citep{h89} (this is $\approx 3.5\%$
of the 87 objects of the sample). We checked whether other BIRG
galaxies matched the Hickson criteria, but could not find any.
BIRGs appear to be preferentially members of groups, although of
groups that are looser than Hickson's compact groups.

\subsection{Interaction Strength and Infrared Emission}

\paragraph{FIR Luminosity \label{corr}}

Is FIR emission directly dependent on -- or even proportional to --
interaction strength?  Our BIRG sample spans a limited range in
\lfir, $10^{10} - 10^{11}$ \lsol. In addition, several objects have
companions whose angular separation is less than half the maximum
width of the IRAS aperture. This implies that a biased correlation
could arise just because, in the closest pairs, we are measuring the
flux of two galaxies. Indeed, if \lfir\ of all small separation
($\simlt$ 1') systems  is treated as an upper limit, there is no
significant correlation between projected linear separation $d_P$\
and \lfir\ (and the correlation is significant if upper limits are
treated as detections!).

A significant correlation appears only if a wider range of \lfir\
is considered. We  added to the BIRG sample data from three
samples for which environmental data are available. We did not
consider systems with a companion whose diameter was $5 Kpc \simlt
D_C \simlt 15$ Kpc, since Fig. \ref{fig02} shows that most of them
may be optical companions. The samples are:

\begin{itemize}
\item Our control sample.
\item The sample of ``very luminous'' IR galaxies by \citet{wu98},
defined as galaxies with $\log$ \lir\ $\simgt 11.15$ in solar
units.
\item The sample of LIG and ULIG  selected by \citet{ssi99},
which turn out to be composed of early and late stage mergers.
\end{itemize}

\begin{figure}

\plotone{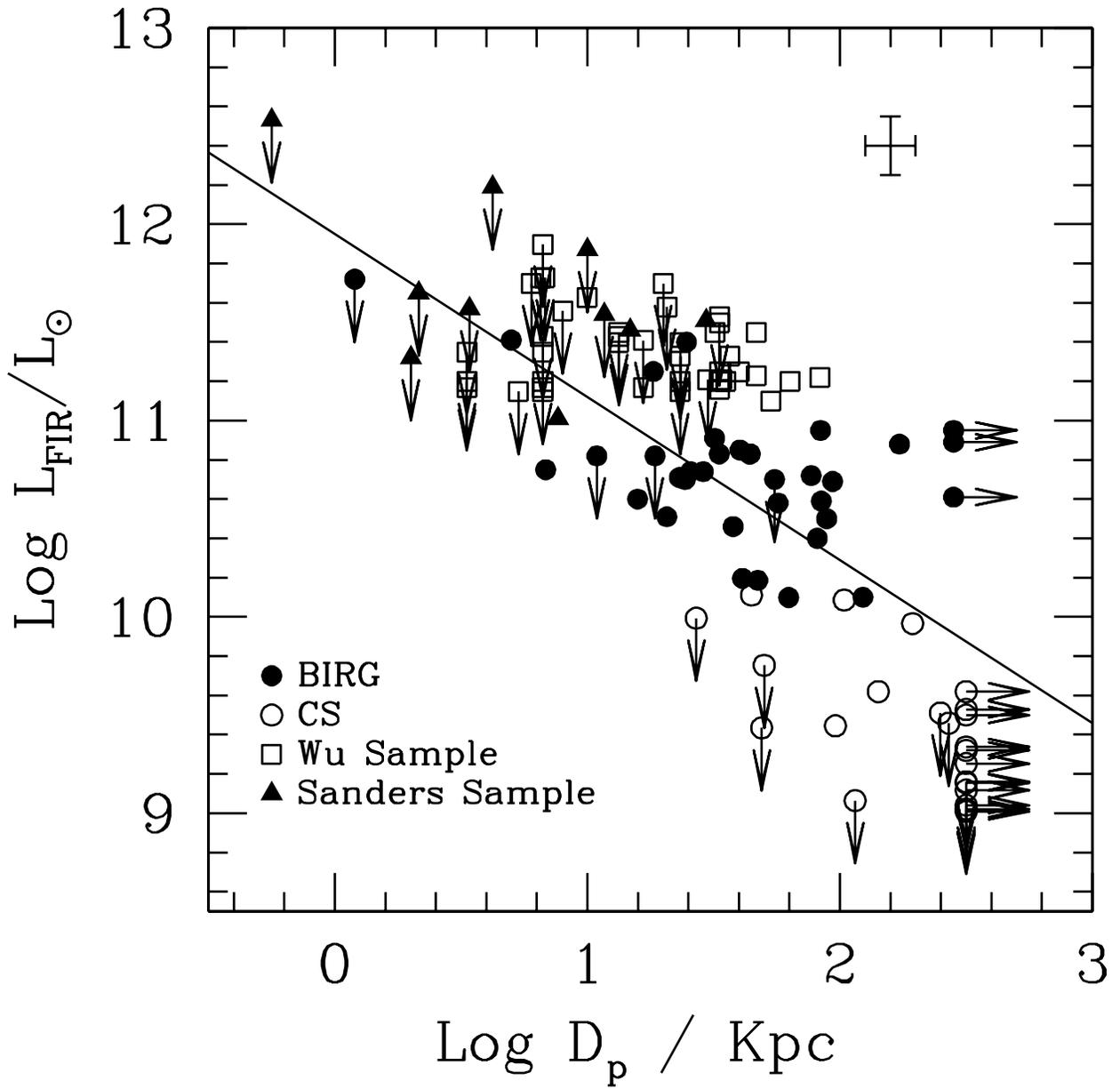}

\caption[f4.eps]{ Far Infrared luminosity L$_{FIR}$\ vs. projected separation $d_p$ for \citet{wu98},
\citet{ssi99}, BIRG, and CS galaxies with bright companion. 107 objects in total. The solid line
corresponds to the best fit. \label{fig04}}
\end{figure}

Fig. \ref{fig04} shows \lfir\ versus $d_P$\ for the galaxies of
the above samples. One has to consider three major limits to the
data:
\begin{itemize}
\item For several objects, the aperture of the IRAS detectors was
larger than the separation between IRAS galaxy and nearest
companion, making impossible to exclude a contribution of the
companion to the measured \lfir.

\item The search radius on the DSS-II was limited to 250 Kpc. There
are some objects (``isolated'') for which there is no companion of
diameter larger than 5 Kpc within this search radius.

\item For several CS galaxies, only upper limits to the fluxes are
set. FIR fluxes were not available for 11 of 22 galaxies (either
isolated or with companion of $D_c \simgt 15 $ Kpc).

\end{itemize}

All of these limitations introduce a censoring on our data. We
considered an upper limit to the $L_{FIR}$\ of the 11 CS objects
that were not detected. For these objects we take the flux density
limits in the four IRAS bands as upper limits to the source flux
density. ``Isolated objects'' were treated as censored in projected
separation of the first companion, and a lower limit to $d_p$\ was
set at our search radius of 250 Kpc (``isolated'' sources are the
ones labelled with horizontal arrows in the lower right side of Fig.
\ref{fig04}). Of course, $d_P$\ values are lower limits to the true
linear separation, which would be the most meaningful parameter to
be correlated. However, the effect of chance projection is to spread
horizontally the points toward the left in the diagram of Fig.
\ref{fig04}, but does not go in the sense of creating a false
correlation. For small $d_P$\ objects, the infrared flux is measured
for the system, and the flux for each galaxy is not available.


We therefore considered \lfir\ for small separation objects as an upper limit. We then applied the
generalized Spearman rank correlation test with inclusion of censored data. The correlation appears to be
statistically significant, with a correlation coefficient $\simlt -0.4$ considering small separation
objects as upper limits. The probability of  the correlation being a chance correlation is $\simlt
10^{-5}$. A best fit using Schmitt's binning regression method yields the relationship $\log L_{\rm FIR}
/L_\odot \approx $ [$-0.83\pm0.21$] $\log [d_P/(1 Kpc)] + [11.95\pm0.33]$\ (see Fig. \ref{fig04}). This
result is confirmed by the presence of an analogous correlation between $d_P$\ and the specific luminosity
at 60 $\mu$m (plot not shown).

\paragraph{FIR colors}

It is important to compare the IR properties of the galaxies with
different strengths of interaction (and thus, with different
projected separation $d_P$). To allow significant results to
emerge in spite of the bias introduced by projection effects, we
considered four interaction classes: (1) mergers (5 Sanders
objects + mergers from BIRG), (2) strongly interacting systems
(galaxies with companion closer than 30 Kpc ($\log d_p <$ 1.5,
where $d_p$ is in Kpc), 5 Sanders objects + BIRG), (3) weakly
interacting systems (galaxies with a companion beyond 30 Kpc
($\log d_p >$ 1.5)), and (4) isolated objects (objects without a
companion within our search radius of 250 Kpc). The objects were
split between BIRG  and control sample of different interaction
classes.

\begin{figure}

\plotone{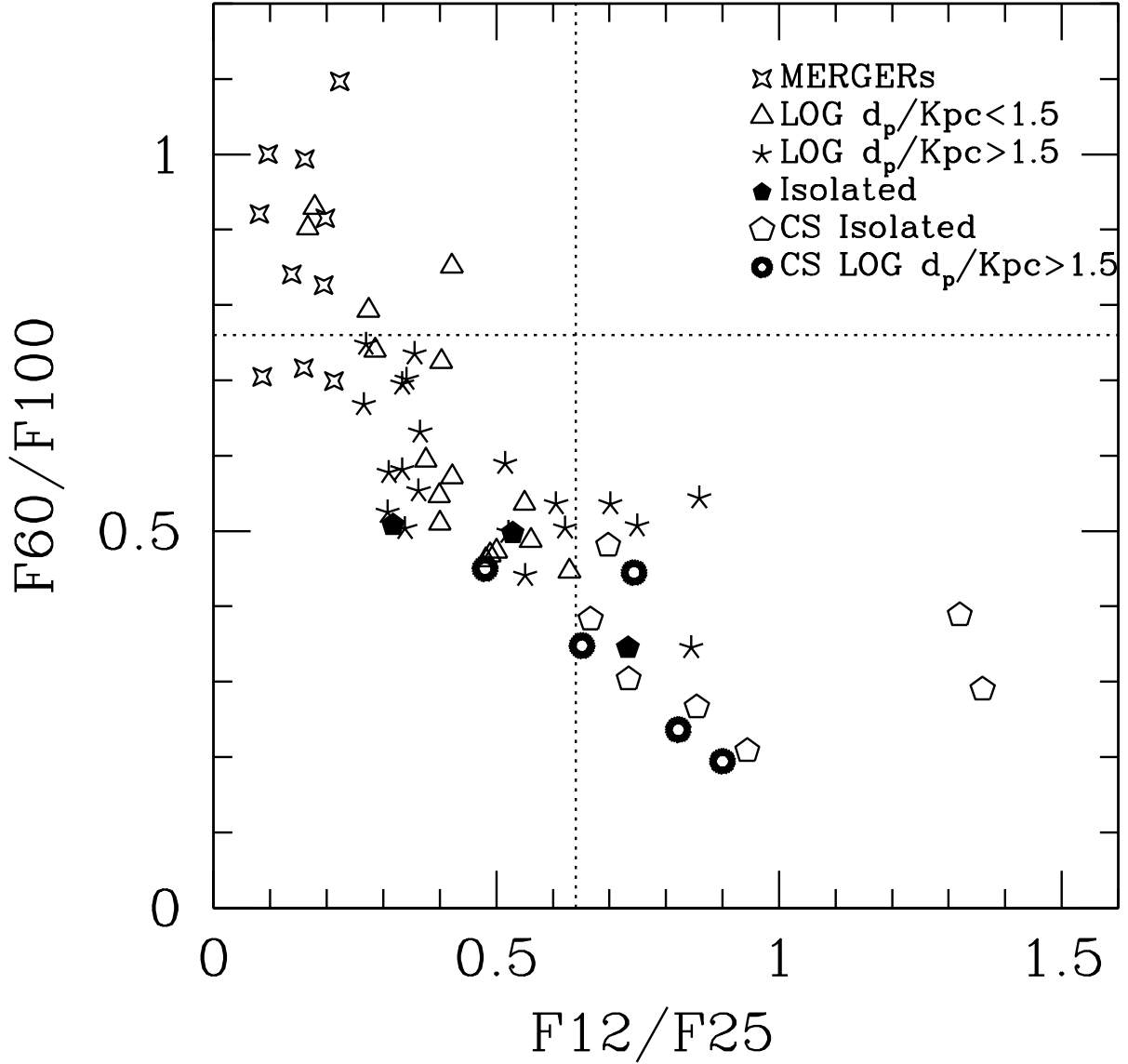}

\caption[f5.eps]{Color-color plot for objects in different interaction classes, and splitted between BIRGs
and CS. The figure is divided in three regions. In the first one (F60/F100$\simgt$ 0.75 and
F12/F25$\simlt$ 0.65), almost all objects are mergers and strongly interacting. In the second region
(F60/F100$\simlt$ 0.75 and F12/F25$\simlt$ 0.65), there is an agglomeration of objects of all interaction
classes. The third region (F60/F100$\simlt$ 0.75 and F12/F25$\simgt$ 0.65) shows only objects with a
companion beyond 30 Kpc, and isolated galaxies.

\label{fig05}}
\end{figure}

Fig. \ref{fig05} shows the F(60$\mu$m)/F(100$\mu$m) vs. F(12 $\mu$m)/F(25 $\mu$m) color-color diagram for
the four interaction classes. Mergers and strongly interacting systems show higher values of F(60$\mu$m)/
F(100$\mu$m) and lower values of F(12 $\mu$m)/F(25 $\mu$m) while isolated objects show lower values of
F(60$\mu$m)/F(100$\mu$m) and higher values of F(12 $\mu$m)/F(25 $\mu$m). Fig. \ref{fig05} is divided in
three regions. In the first one (F(60$\mu$m)/F(100$\mu$m) $\simgt$ 0.75 and F12/F25$\simlt$ 0.65), almost
all objects are mergers and strongly interacting. In the second region (F(60$\mu$m)/F(100$\mu$m)$\simlt$
0.75 and F(12 $\mu$m)/F(25 $\mu$m)$\simlt$ 0.65), there is an agglomeration of objects of all interaction
classes. However the 3 mergers in this region are near the border to the first region, and their IR colors
are very close to the values of the first region mergers. The third region
(F(60$\mu$m)/F(100$\mu$m)$\simlt$ 0.75 and F(12 $\mu$m)/F(25 $\mu$m)$\simgt$ 0.65) shows only objects with
a companion beyond 30 Kpc, and isolated galaxies.


\paragraph{Overall Properties}
Table 2 reports average and sample standard deviation values of the
parameters considered in our analysis (Column 1), for different
interaction strength classes.  Columns 2--5 report sample average
and sample standard deviation for CS and BIRG isolated galaxies
($d_P \simgt 250$ Kpc). Columns 6--9 report values for BIRG and CS
weekly interacting galaxies with $d_P \simgt 30$ Kpc. The next
columns list the sample average and standard deviation for the BIRG
sample for the remaining two interaction classes: strongly
interacting, and mergers (there are no CS galaxies with $d_P \simlt
30$ Kpc nor mergers). The last four rows provide standard estimates
of star-formation-related parameters: (1) the Star Formation Rate
(SFR), which was computed from \lfir\ using the standard
relationship $SFR \approx 4.5\times 10^{-44}L_{FIR, ergs s^{-1} }$
M$_{\odot}$ yr$^{-1}$ \citep{k98}; (2) hydrogen molecular mass \m2\
(collected from various sources in literature and available for 41
objects); (3) the ratio \lfir/\m2; (4) the depletion time in yr
simply defined as the hydrogen molecular gas mass assumed over the
SFR, $\tau_{H_2} = $ \m2\ / SFR.

There is a clear continuity on FIR properties and SFR from isolated
objects to mergers (except for the 3 isolated BIRGs, but see below).
\lfir\ increases with the interaction strength as indicated by our
correlation analysis. Systematic differences in FIR  color are also
appreciable.  The depletion time  is $\simlt 10^9$ yr for {\em all}
interaction classes (including isolated objects) in the BIRG sample.
In the CS objects, $\tau_{H_2} \sim 10^{10}$ yr, comparable to the
Hubble time. There is a monotonic trend from isolated galaxies to
mergers, in terms of increasing SFR and decreasing $\tau_{H_2}$, but
it is noteworthy that \m2\ is not statistically different in the
various interaction classes.

Isolated objects from the control sample, and isolated objects
from the BIRG sample, have impressively different \lfir. This
apparent contradiction needs an explanation. There are only 3
isolated BIRGs. NGC 5937, NGC 7083, and NGC 5936 did not show a
companion larger than 5 Kpc on the DSS-II within 250 Kpc.
However, all of these galaxies present peculiarities. (1) NGC
5937 has a distorted morphology, and it may have a loop of gas
which could be a signature of interaction. (2) NGC 7083 is a
barred Sc galaxy that hosts a LINER. It looks perturbed because
of an off-centered loop. (3) NGC 5936 has a highly distorted
morphology, which may be indicative of recent interaction. These
galaxies may have been disturbed by the presence of a small
companion disrupted or projected over the main galaxy. Isolated
CS galaxies do not show distortions or peculiarities that could
make them special objects in terms of morphology or interaction.

\section{Discussion \label{disc}}

The percentage of companion galaxies within $3D_S$\ and the
  distributions of observed and physical companions show an
highly significant excess for the BIRGs.  The difference between
BIRGs and CS galaxies is especially striking if large companions
with $D_C \simgt 20$ Kpc are considered (the BIRG galaxies have 3--4
times more companions within $\approx$ 140 Kpc; strongly interacting
systems in the CS may be $\simlt$ 1\%). Our results also indicate a
direct relationship between interaction and enhancement of IR
emission. We have considered a very large range in \lfir, $ \sim
10^{8.5} - 10^{12.5}$ \lsol, which is unprecedented and probably
sufficient to overcome the bias introduced by random projection of
separation. This may explain why, with some notable exceptions  (e.
g. \citet{sm96} and references therein), several previous analyses
did not found any convincing correlation between $d_P$\ and \lfir\
among interacting galaxies. Our result extends to a lower \lfir\
range and quantifies results that were known qualitatively for LIRGs
and ULIRGs (\citet{ssi99}).

\subsection{Implications for Star Formation}

An increase in \lfir\ can be observed across a sequence from
isolated galaxies to strongly interacting systems. Color
variations are consistent with the emergence of a FIR continuum
component whose luminosity and colors are correlated. This
component can be associated to thermal re-radiation by dust of
hot stars continuum emission. In the most extreme cases of
isolated CS galaxies, we may have only a cold cirrus component,
T$\sim 20^\circ$ K. At the other end of the FIR color-color
diagram, a ``hot'' component peaking around 100-60 $\mu$m may
have become prominent. The increase in \lfir\ can be largely
ascribed to an increase in the SFR, as shown in many previous
studies (\citet{k98}, and references therein; \citet{st92}). For
the ``hottest'' sources (F(25$\mu$m)/F(60$\mu$m)$\simgt$ 0.2),
however, the reprocessed continuum  may be due to a non-thermal
source \citep{dg92}.

The difference in \lfir\ and \lfir/\m2\ (a factor of more than 100
from mergers to isolated CS objects, see Table 2) suggests that
strong interactions ($d_P \simlt 30$ Kpc) are a necessary and
sufficient condition for an extreme SFR and for a ``starburst''
(defined as star formation that cannot be maintained over the Hubble
time), at least for the galaxies of our sample (this result may not
be generally true if  not all mergers of gas rich-galaxies are
infrared luminous) . A companion that has approached to less than 30
Kpc to a galaxy may need a time $\simgt 3\times 10^8 d_{30Kpc}
v^{-1}_{100 Km s^{-1}}$ yr to move beyond this distance. The mean
depletion time for strongly interacting galaxies is $\approx
5\times10^8$ yr (Table 2). In this case, the interaction time and
the $\tau_H$\ are comparable. This means that a galaxy may exhaust
its gas before an interaction episode is over, on a time much less
than the Hubble time.

On the other hand, the SFRs of  weakly interacting galaxies (CS
galaxies with $d_P \simgt$ 30 Kpc) do not show values that may be
considered extraordinary  (SFR $\approx 0.52 M_\odot/yr$).  For
objects whose companion is separated by $d_P \simgt 30 $ Kpc, the
average $d_P$\ is  approximately 112 Kpc and 67 Kpc in the CS and in
the BIRG sample respectively. The SFR is $\approx$ 10 times larger
in the BIRG than in the CS. This is consistent with tidal forces
($\propto d_P^{-3}$) driving the SFR increase. A weak interaction
may produce a moderate enhancement of the SFR of a galaxy, but not
lead to dramatic effects on its secular evolution. An important
implication of our results is that at least part of the large
dispersion (a factor $\sim$10) for the SFR in galaxies of a
particular morphological type (see \citet{k98}) may be explained by
weak interactions (cf \citet{ht01}).

\subsection{Relationship between Star Forming and Seyfert Galaxies \label{seyf}}

Our work and many previous ones leave no doubt that gravitational
interaction leads to an increase of the SFR in gas-rich galaxies.
Less clear is the relationship between interaction and the
occurrence of non-thermal nuclear activity. In the simplest scheme
of Seyfert unification, Seyfert 1 and Seyfert 2 are different
because of orientation (see e.g., \citet{a93} for a review): a
molecular torus makes obscuration a major factor in the appearance
of an active nucleus. However, interaction may be a factor leading
to the formation of the obscuring torus itself, and to the
production of extensive circum-nuclear star formation. A
significant role of interaction introduces an additional degree
of freedom \citep{dd99a} related to environment and, in a broad
sense, evolution.

\subsubsection{The Environment of Seyfert Galaxies}

The main question is then, what is the environment of Seyfert
Galaxies? The most recent works have found a positive excess of
large companions among Seyfert 2s (Sy2s), but not among Seyfert 1s
(Sy1s) \citet{dd99a}; \citet{ls95}; \citet{dr98}). This challenges
previous results suggesting an excess without differences between
Sy1s and Sy2s (\citet{d84}; \citet{r95}). Problems here may arise
because of intrinsic inhomogeneity in the discovery techniques of
Sy2 galaxies, as discussed explicitly by \citet{m91}. In addition,
it has to be taken into account that discovery methods for Sy2
galaxies like the UV excess and the FIR color are sensitive to
enhanced star formation. For instance, \citet{schmitt01} selected a
FIR-flux limited sample on the basis of FIR color. They found that
31\%$\pm$10\%\ of Sy1s and 28 \%$\pm$7\%\ of Sy2s have companions
(optical + physical) within three diameters. These frequencies are
very similar to the frequency found for Sy2s by \citet{dd99a}
(companion diameter $\geq 10$Kpc within 60 Kpc, the case most
similar to the one considered by \citet{schmitt01}). By introducing
a bias in favor of star-forming Sy1s, their selection criterion may
have simply {\em increased} the fraction of interacting Sy1s
(\citet{dd99a} found 21\%!) with respect to Seyferts 2. As it can be
seen below, almost all AGN from the BIRG sample show evidence of
significant star formation {\em and} belong to interacting systems.

Since an excess is found from uniformly-distributed samples
\citep{dd99a} and also for a Seyfert sample selected from the Center
for Astrophysics redshift survey (\citet{dr98}), we consider {\em an
excess of bright companion among Seyfert 2 galaxies, and no excess
among Seyfert 1s with respect to a suitably-chosen control sample of
non-active galaxies} as the most accurate representation of the
Seyfert environment.

\subsection{Star Forming \& Seyfert Galaxies: an Evolutionary Sequence?}

The  result of this paper which is relevant at this point is that
BIRGs seem to have, to a high confidence level, more large and
close companions ($D\geq 10$ Kpc, $d\leq 60Kpc$) than Seyfert 1s,
and seem to be similar in their environment to Seyfert 2s (cf.
\citet{dd99a}). This statistical result gives support to a scheme
that several workers have considered (\citet{he89}, and
references therein; \citet{s88}). The scheme is an evolutionary
sequence for AGN driven by interaction:

\begin{equation}
Interaction \Rightarrow Starburst \rightarrow
\underbrace{Seyfert~2 \rightarrow Seyfert~1}
\end{equation}

where the brace indicates that Sy1 and Sy2 may be actually the same
kind of objects seen in different orientations. There are several
lines of (admittedly circumstantial) evidence that also support this
simple evolutionary path: first, the contribution of thermal
emission to the bolometric luminosity appears to decrease along the
sequence (\citet{dd96}). Sy1 nuclei have been revealed in several
evolved mergers (for instance, see \citet{raf93}). Second, there are
several active galaxies in the BIRG sample. Of 87 galaxies, 17\%\
host a Seyfert 2 nucleus (15 objects), but only 2.5\%\ host a
Seyfert 1 nucleus (2 objects). There is no statistical difference
between the \lfir\ of active and non-active galaxies, except for a
slightly higher value in the Sy1 objects. The value of
F(25$\mu$m)/F(60$\mu$m) for Sy2s is $\approx 0.18$ and for Sy1s
$\approx 0.20$, compared to$\approx 0.13$ for non-active galaxies.
F(25$\mu$m)/F(60$\mu$m) is larger in Sy1s and Sy2s due to the
contribution to the continuum of a non-thermal source
(\citet{dg92}). Estimating in a careful way the ratio
thermal/nonthermal emission for the BIRGs is not possible from
published data. However, 73\%\ of the Seyfert 2 (11 of 15) show
evidence of significant star formation (there is evidence of a
circum-nuclear starburst in 45\%\ of the star forming Sy2s). Also the
2 BIRG Seyfert 1 galaxies  show evidence of a circum-nuclear
starburst (this makes selecting samples of Seyfert 1 and Seyfert 2
from \lfir\ even more improper for environmental studies than
selecting them from catalogues!)

The evolutionary sequence outlined above can be understood in
three different ways.

\begin{enumerate}

\item  It can be read as a sequence of
obscuration properties: (a) fully-obscured Seyfert 1s (i.e., seen
as a Seyfert 2 from all viewing angles), (b) obscuration
dependent on viewing angle Sy1s (the ``unification'' Sy1 and Sy2
scenario), (c) almost fully unobscured Seyfert 1s.

\item  It can be a sequence of AGN power; a possibility is that the accretion rate
may be insufficient to maintain a Broad Line Region (BLR) in some
Sy2s.

\item A low power may also occur in
its earlier stages just because the central black hole is rather
of low mass, maybe because the black hole {\em was not originally
present}.

\end{enumerate}

A wealth of X-ray data show that most Seyfert 2 are consistent with an AGN X-ray spectrum increasingly
less absorbed at energies $\simgt 5$ KeV. This means that an AGN has been already switched on in many, if
not all, Seyfert 2 galaxies \citep{moran01,matt97}. This results supports the obscuration sequence (since
the power of the AGN will be roughly the same in different types). In this scheme, Seyfert 2s may appear
as the low-luminosity analogues of ULIRGs, which have been suggested to be precursors of quasars.

\citet{t01} studied a sample of Seyfert 2 galaxies to determine
how many of them were obscured Sy1s i.e., showed a hidden broad
line region (HBLR) in polarized light. He concluded that non-HBLR
Sy2s are not more obscured than HBLR Sy2s, but less powerful AGN.
This result goes against the obscuration scenario, and favors the
AGN power scenario (points 2 and 3). \citet{gu01} studied the
properties of 51 Sy2s with evidence of high circumnuclear SFR.
They found that while Sy2s with a HBLR have similar
Infrared-Radio properties as Sy1s, Sy2s without a HBLR have
properties similar to Starbursts. These results can be
straightforwardly understood in the context of an evolutionary
scheme. While objects without a HBLR are ``younger" Sy1s (whether
very obscured or with very low AGN power), Sy2s with a HBLR are
``young" Sy1s that may keep forming stars on their nuclear region
but that are less obscured or with higher AGN power.

Obscuration, low accretion or small black hole mass could be
therefore the main physical factors behind any evolutionary
sequence. However, we think that there is presently  not enough
evidence to decide in favor of one of these factors.

\subsubsection{Environmental Effects as Drivers of any Evolutionary Sequence}

The time needed for Sy1s to emerge (whether as unobscured or as
high power AGN) could be longer than the escape time of an
unbound companion from the very close environment, or comparable
to the time-scale needed for an evolved merger ($\sim 10^9$ yr).
This will naturally explain why Starbursts and Sy2s are found
more often with closer companions.

For AGN triggering in a gas rich galaxy, the occurrence of a tidal
perturbation may be more relevant than its duration
\citep{keel96}. An hyperbolic encounter may well trigger a radial
flow in the innermost regions of a gas-rich galaxy. The time
needed by the companion to move away by 30 Kpc is $\sim 1.\times
10^8 d_{30Kpc} \Delta v^{-1}_{300 Km s^{-1}}$ yr. The timescale
for a clump of gas to fall from the outer regions of the nucleus
(a few hundreds of parsecs) to the inner central pc is $\simgt
0.1$ Gyr (\citet{b00}), and this can be considered a lower limit
to the time needed for the onset of the active nucleus.
Therefore, an hyperbolic encounter with moderate $\Delta v_r$\
can be such that the companion escape from the close vicinity
($\approx 60 $ Kpc) of the Seyfert galaxy, leaving a
non-interacting Seyfert 1 nucleus. If obscuration is significant,
or if the AGN power is small  (because of low accretion rate, or
of an undermassive  central black hole), then a longer timescale
may be necessary before a Seyfert 1 nucleus is actually detected.
While BIRGs and Sy2 galaxies have richer environments than Sy1s
at distances $\simlt 60 Kpc$, the cumulative distribution of the
projected separation for the first companion (Fig. \ref{fig03})
shows that the environmental difference for Sy1s, with respect to
Sy2s and BIRGs decreases dramatically beyond $\approx$ 120 Kpc.
This means that, while Sy2s and BIRGs have close companions, Sy1s
do have companions, but at higher distances ($d_p \simgt 100$
Kpc). Sy1s do not show close companions simply because any
activity-triggering interaction  took place in the past, and, on
average, Sy1 galaxies would not be considered interacting
following our statistical criterions.

The limitations of our analysis regarding small companion galaxies
($D_C \simlt 10 Kpc$)  leave open other main possibility  to
account for type-1 activity. It has been proposed that Sy1s may
be the result of a ``minor merger" which purportedly may not lead
to a dramatic star formation close to the center of the galaxies
and hence to heavy obscuration (\citet{dr98},\citet{t99}). N-body
simulations of minor mergers show that they produce disturbances
in the morphology of the larger galaxy in the first Gyr of the
onset of the merger, but do not destroy the galactic disk
(\citet{w96}). \citet{c00} did not found higher levels of
asymmetry in Seyfert galaxies than in normal galaxies (in
agreement with our work, he found that the most asymmetric
galaxies were interacting systems with HII-like spectra). He
concluded that, if minor mergers trigger AGNs, they appear to do
so only in the late stages of the mergers ($\sim$ 1 Gyr after the
merger onset). Minor mergers also boost the star formation of the
larger galaxy, but this process is not necessarily very dramatic
(the induced SFR may be as low as $\approx 2 M_{\odot}/yr$)
especially after the first $0.5$ Gyr (\citet{r00}).

The previous mechanisms suggest a revision that complement the
unification scheme for Seyfert galaxies, and favor the idea of a
long timescale to let type-1 AGN emerge. It is interesting to
stress that times for the onset of this kind of activity are in
agreement with the time needed to let any unbound companion fly
at least few tens of Kpc, or to have a full or a ``minor'' merger
($\sim 1$ Gyr).

\subsubsection{The Unlikely Alternative: No Effect on AGN Triggering by
Tidal Forces}

An alternative interpretation for the environmental results for BIRGs, Sy2s, and Sy1s is that interactions
may trigger only high SFR but no nuclear, non-thermal activity. Seyfert 2s may show an interaction-induced
enhancement in the SFR as do any other interacting galaxies, at least on average. This implies  two
populations of Sy2 galaxies \citep{sb01}:
\begin{itemize}

\item  interacting Sy2s with  high SFR. The morphology of these
galaxies should be distorted due to the interaction, and the
interaction could be responsible for the obscuring torus of dust,
if it exist. The properties of this Sy2 galaxies should be
similar to those of star forming galaxies.

\item Sy2s isolated and without any circumnuclear starburst.
Due to the lack of interactions, the morphology of these objects
should be very symmetric, without distortions and, as Sy1s, these
galaxies should not have any excess of companions when compared to
normal galaxies.
\end{itemize}

If this distinction is correct, strong interactions and
non-thermal activity could be fully unrelated phenomena
(\citet{c00}). The issue would suffer a 30-year setback.  The
crucial test is then whether the excess of interacting Seyfert  2
with respect to normal galaxies is real for a {\em complete sample
of Seyfert 2}. If not, then there would be no support for a
relationship between interaction and Seyfert type activity. If
yes, then the evolutionary sequence above may be appropriate.
Defining a complete sample of Seyfert 2 galaxies is tricky, but,
as already noted, the results based on a limited CfA sample
suggest that the difference between Seyfert 2 and Seyfert 1 may
not be due to sample selection biases. As noted earlier, the value
of F(25$\mu$m)/F(60$\mu$m) for Sy2s is $\approx 0.18$ and for Sy1s
$\approx 0.20$. This result argues against  two Sy2 populations,
since Sy2s interacting and with high SFR (like the Sy2s of the
BIRG sample) should have properties more similar to star forming
galaxies, rather than to to Sy1s (and thus, lower
F(25$\mu$m)/F(60$\mu$m) ratios).

\section{Conclusions}

We studied the environment of Bright IRAS galaxies and compared it
to the one of FIR low emitters, as well as to the one of Seyfert 1
and 2 galaxies. We found that, on average, BIRGs galaxies are more
often in interaction, and that their ``interaction strength'' is
higher than in a sample of optically-selected galaxies. Our
results show a weak anticorrelation between the projected
separation of the first companion and the FIR luminosity of a
galaxy, which means an anticorrelation between $d_P$ and the SFR.
This extends previous results for luminous and ultra-luminous FIR
galaxies. The FIR properties show a clear and smooth continuity
as a function of interaction strength, going from very low FIR
activity in isolated normal galaxies to very high activity in
mergers. A consequence is that the FIR luminosity function as a
function of morphological types is meaningful only for strictly
isolated, unperturbed systems. The similar environment found for
Seyfert 2 and BIRG galaxies supports the possibility of an
evolutionary link between Starburst, Seyfert 2 and Seyfert 1.

\acknowledgements

This research has made use of the Digitized Sky Surveys which
were produced at the Space Telescope Science Institute, under
U.S. Government grant NAG W-2166, and of the NASA/IPAC
Extragalactic Database (NED), which is operated by the Jet
Propulsion Laboratory, California Institute of Technology, under
contract with the National Aeronautics and Space Administration.
P.M. acknowledges  financial support from grant Cofin
2000-02-007. Y.K. acknowledges financial support from grant
PAEP201304. D.D.H. acknowledges financial support from grant
IN115599 PAPIIT DGAPA-UNAM.

\clearpage

\begin{deluxetable}{ccccccc}
\tablecolumns{6} \tablewidth{0pc} \tablecaption{  Fraction of
observed, optical, and physical companions}
\tablehead{\colhead{Sample Id.} & \colhead{Sample Size} &
\multicolumn{3}{c}{Frequency of Companions (\%)} &
\colhead{Significance\tablenotemark{a}}
\\ \cline{3-5}
& & \colhead{ Observed} & \colhead{Expected} & \colhead{
Physical} & \% } \startdata \cutinhead{Companion Diameter $\geq
5$ Kpc}

BIRGs         & 87                    & 40.3\%          & 36.3\%
& 4\%  &
\nodata  \\
CS  & 90                  & 42.6\%          & 36.1\%  &  6.5\% &  not signif. \\
\cutinhead{Companion Diameter $\geq 10$ Kpc}
BIRGs   & 87             & 58.4\%          & 20\%     & 38.4\%  & \nodata  \\
CS    & 90              & 29\%          & 18.4\%  &  10.9\% & 99.9\%     \\
\enddata
\tablenotetext{a}{Statistical significance for the hypothesis that
the listed samples are different from the BIRG sample.}
\end{deluxetable}

\clearpage

\begin{deluxetable}{cccccccccccccc}
\rotate \tabletypesize{\scriptsize}
 \tablecolumns{13}
\tablewidth{0pc} \tablecaption{  FIR properties for galaxies with
different interaction strengths}

\tablehead{\colhead{Parameter} & \multicolumn{4}{c}{ISOLATED} &
\multicolumn{4}{c}{Separation $> 30 Kpc$} &
\multicolumn{2}{c}{Separation $< 30 Kpc$} &
\multicolumn{2}{c}{MERGERs} &
\\ \cline{2-5} \cline{6-9} \cline{10-11} \cline{12-13}

 & \colhead{CS} & \colhead{St. Dev} & \colhead{BIRGs} &
\colhead{St. Dev} & \colhead{CS}  & \colhead{St. Dev} &
\colhead{BIRGs} & \colhead{St. Dev} & \colhead{BIRGs } &
\colhead{St. Dev} & \colhead{MERGERs} & \colhead{St. Dev}  }

\startdata

Number of objects & 11 &\nodata &  3 &\nodata & 11 &\nodata & 20 & \nodata & 16 &\nodata & 10 & \nodata \\

$<d_p>$ \ (Kpc) &  $>250$  &  \nodata &   $>250$   & \nodata  &
$111.73$ & $71.50$ &
$66.94$ & $35.05$  & $18.20$ &  $7.42$  &  $1.81$  &  $1.17$ \\

$<L_{FIR} >$ \ ($10^{10}L_{\odot}$)   &  $0.21$  & $0.13$ &
$6.92$ & $2.53$ & $0.92$ & $0.47$ & $3.83$ & $2.30$ & $16.8$ &
$18.8$ &
$71.6$ & $103.2$ \\

$<L_{12\mu}>$($10^{30}$$ergs^{-1}Hz^{-1}$) & $0.27$ & $0.15$ &
$1.35$ & $0.20$ & $0.34$ & $0.21$ & $1.19$ & $0.71$ & $5.16$ &
$8.76$ & $9.43$ & $1.99$
\\

$<L_{25\mu}>$($10^{30}$$ergs^{-1}Hz^{-1}$) & 0.29 & 0.12 & 2.98 &
1.71 & 0.39 & 0.30 &  2.61 &  1.44 & 14.6 & 22.2 & 48.6 & 88.4 \\

$<L_{60\mu}>$($10^{30}$$ergs^{-1}Hz^{-1}$) & 1.02 & 0.51 & 19.4 &
8.70 &    3.38 &    2.80 & 20.2 &    10.5 & 70.3 & 67.7 & 275.1 &
374.2 \\

$<L_{100\mu}>$($10^{30}$$ergs^{-1}Hz^{-1}$) &   3.40 & 2.27 & 42.3
&    13.1 &    9.85 &    6.84 & 37.1 & 22.0 & 103.2 & 79.1 &
285.3 & 344.4 \\

$<F12/F25>$ & 0.94 &   0.29 &   0.53  &  0.21  &  0.72  &  0.16 &
0.48 &  0.19 &  0.41  &  0.13  &  0.16  &  0.05 \\

$<F60/F100>$  &  0.33  &  0.09  &  0.45  &  0.09  &  0.34  &  0.12
& 0.57  & 0.10  &  0.63  &  0.17  &  0.87  &  0.14 \\

$<M_(H_{2})>$ \ ($10^9$ $M_\odot$) &  3.77  &   1.68  &  2.95 &
not avail. & 7.12 & 2.11 & 6.40 &  5.89 &  7.55  & 6.81 & 9.60 &
9.04
\\

$<Lfir/M_(H_{2})>$ \ ($L_\odot/M_\odot$)  & 0.52  &  0.42 & 26.30
& not avail. & 0.85  & 0.53  &  13.56 & 12.65 &  31.47 & 35.94 &
118.0  & 105.77 \\

$<SFR>$\tablenotemark{a} \ ($M_\odot$ $yr_{-1})$ & 0.36  &  0.23 &
12.10 &  4.42 &
1.45 & 1.15 & 6.71 & 4.03 & 29.56 &  32.95  & 125.34 & 180.09 \\

$<T_{H_2}>$\tablenotemark{b} \ ($10^9$ $yr$) & 19.6 &    17.9 &
0.22 & not avail. &
8.39 & 5.22 & 0.91 &  0.86 &    0.51 &    0.59 & 0.09 & 0.08 \\

\enddata
\tablenotetext{a}{$SFR \approx 4.5\times 10^{-44}L_{FIR, ergs
s^{-1} }$ M$_{\odot}$ yr$^{-1}$ \citep{k98}}
\tablenotetext{b}{$\tau_H = $ \m2\ / SFR.}
\end{deluxetable}

\clearpage

\end{document}